\documentclass[useAMS,usenatbib]{mn2e}
\usepackage{epsfig,subfigure,amsmath}
\usepackage{color}

\definecolor{purple}{rgb}{1,0,1}

\bibpunct[; ]{(}{)}{;}{a}{}{,}

\title[A response to arXiv:1310.2791]{A response to arXiv:1310.2791: ``A self-consistent public catalogue of voids and superclusters in the SDSS Data Release 7 galaxy surveys''}
\author[P.~M. Sutter et al.]
{
\parbox{\textwidth}{
{P.~M. Sutter}$^{1,2,3,4}$ \thanks{Email: sutter@iap.fr},
Guilhem Lavaux$^{3,5,6,7}$,
Benjamin D. Wandelt$^{1,2,5,8}$, and
David H. Weinberg$^{3,9}$
}
\vspace{0.4cm}\\
\parbox[c]{\textwidth}{
$^{1}$ UPMC Univ Paris 06, UMR7095, Institut d'Astrophysique de Paris, F-75014, Paris, France \\
$^{2}$ CNRS, UMR7095, Institut d'Astrophysique de Paris, F-75014, Paris, France \\
$^{3}$ Center for Cosmology and Astro-Particle Physics, Ohio State University, Columbus, OH 43210\\
$^{4}$ Department of Physics, University of Illinois at Urbana-Champaign, Urbana, IL 61801\\
$^{5}$ Department of Physics \& Astronomy, University of Waterloo, Waterloo,
ON,  N2L 3G1 Canada \\
$^{6}$ Perimeter Institute for Theoretical Physics,
Waterloo, ON, N2L 2Y5, Canada \\
$^{7}$ Canadian Institute for Theoretical Astrophysics, 60 St. George St.,
Toronto, ON M5S 3H8 Canada \\
$^{8}$ Department of Astronomy, University of Illinois at Urbana-Champaign, Urbana, IL 61801\\
$^{9}$ Department of Astronomy, Ohio State University, Columbus, OH 43210\\
}}

\begin{document}

\maketitle

\label{firstpage}

\begin{abstract}
Recently,~\citet{Nadathur2013} 
submitted a paper discussing a new cosmic void catalog. 
This paper includes claims
about the void catalog described in~\citet{Sutter2012b}.
In this note, we respond to those claims, 
clarify some discrepancies between the text of~\citet{Sutter2012b} 
and the most recent version of the catalog, 
and provide some comments on the differences 
between our catalog and that of~\citet{Nadathur2013}. 
All updates and documentation 
for our catalog are available at~{http://www.cosmicvoids.net}.
\end{abstract}

The potential of voids  to constrain cosmology and fundamental
physics is being increasingly recognized~\citep{VandeWey2011b} and it is
therefore very important to have a clear understanding of void
definition used.
In last week's arXiv submission~\citet{Nadathur2013} (hereafter NH13) 
described a different
procedure for building a void catalog based on {\tt ZOBOV}~\citep{Neyrinck2008}.
NH13 contained several implied or explicit criticisms of
the catalog we made available at
{http://www.cosmicvoids.net} (Sutter et al. 2012;~hereafter S12).
We take this opportunity to provide some clarifications
in this comment.

The principal criticism in NH13
of the S12 void catalog
is that the regions identified as voids are not, on the whole, underdense.
This is by design: the {\tt ZOBOV} watershed algorithm 
includes the surrounding high-density walls in the void 
definition~\citep{Neyrinck2008}.
Indeed, Figure 7 of NH13 shows identical behavior in their voids. 
However, the voids in S12 \emph{are} underdense
near the center, as demonstrated by the radial density profiles plotted in
Figure 9 of S12. 

\begin{figure} 
  \centering 
  {\includegraphics[type=png,ext=.png,read=.png,width=\columnwidth]{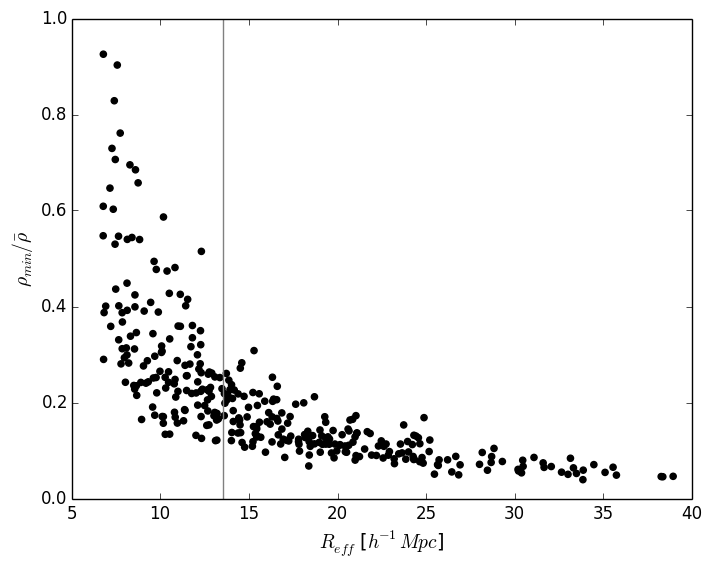}}
  \caption{Minimum zone density in the void versus void effective 
           radius for voids in the \emph{bright1} sample. 
           The vertical line indicates $2 R_{\rm eff, min}$.
          }
\label{fig:rhominvr}
\end{figure}

Another major criticism in
NH13  is that some voids in S12 have high minimum densities 
(and also high mean densities).
When these voids are included,
a stacked radial density profile of all voids shows no clear 
underdensity (Figure 9 of NH13).
However, as Figure~\ref{fig:rhominvr} shows, voids with high 
$\rho_{\rm min}$ (and hence also high $\rho_{\rm void}$) tend to 
have radii near the mean intergalaxy separation $n_g$.
These voids are naturally more sensitive to Poisson
noise fluctuations and less robust in properties than larger voids.
Small voids tend to have higher compensation regions~\citep{Ceccarelli2013, 
Hamaus2013}, leading to 
higher mean densities in the watershed. However, they also have 
\emph{lower} density contrasts~\citep{Sutter2013a}. These effects 
combined will tend to wash out a stacked density profile.

In our catalog we impose a sharp cutoff at $R_{\rm eff} = n_g^{-1/3}$, but
in the catalog documentation we
caution users that voids
near this cutoff may not be reliable
and we recommend that a higher threshold (e.g., $R_{\rm eff} > 2 R_{\rm eff,min}$)
be adopted for most analyses.  In general our approach has been to
provide a catalog with loose cuts but offer advice on targeted cuts that
users may wish to impose for their analyses, while NH13's approach
is to impose what they consider best-practice cuts in the catalog that 
they provide.  Obviously, each approach has its virtues.

Below we respond to the other main points raised by NH13.

\section{Other Responses}

\emph{Survey boundary contamination} - NH13 claim that 30\% of our 
voids have core particles that are adjacent to boundary particles 
(mock particles introduced to handle survey edge effects). 
We cannot replicate this claim. We have modified the original 
{\tt ZOBOV} algorithm to enforce bijectivity in the Voronoi graph and 
use a different splitting technique to minimize problems when 
joining subregions.
Not including these refinements may explain why NH find a different 
result. Also, 
the use of earlier versions of the {\tt qhull} 
library\footnote{http://www.qhull.org}, which {\tt ZOBOV} 
uses to construct the Voronoi graph, can result in different 
tessellations.

\emph{Statistical significance of voids} - NH13 claim that 
most of our voids are indistinguishable from Poisson fluctuations based 
on low density contrasts. 
The voids in NH13 also have very low 
contrasts (their Figure 4), and the cut on contrast ratio 
recommended by~\citet{Neyrinck2008} would 
eliminate nearly all voids. Instead, they cut their sample 
on $\rho_{\rm min}$.  As Figure~\ref{fig:rhominvr} shows, the 
most problematic voids are the ones nearest the 
mean galaxy separation, and 
if these voids were 
to be judged undesirable for a particular application
a simple cut on larger void radii 
would remove most of them.

\emph{Choice of coordinate system} - NH13 claim that our catalog is 
problematic 
because the abundances of voids in redshift space (with galaxy 3-d
positions computed using $D_A(z) = cz$) and comoving space (assigning
positions in comoving coordinates for an assumed $\Lambda$CDM
cosmology) do not match.
 Since we impose a fixed radius cutoff, and 
the mean galaxy separation changes with the coordinate transformation, 
we will naturally get different abundances. While voids in redshift and 
comoving space tend to have different shapes and sizes~\citep{Ryden1996}, 
and the robustness of watershed techniques under coordinate 
transformations needs to be more 
carefully studied,
the analyses of~\citet{Planck2013b},~\citet{Pisani2013}, 
and~\citet{Melchior2013}
indicate that
we still capture physical underdensities. 
Additionally, there is significant correspondence in the 
positions on the sky between the 
the redshift- and comoving-space voids in our catalog.
We identified voids in redshift space for our application of the 
Alcock-Paczynski test~\citep{Sutter2012b}, but we have always provided 
a comoving-space version of the catalog and clearly noted this 
in the documentation. Note that in both coordinate systems we do not 
attempt to remove peculiar velocities.

\section{Discrepancies between S12 text \& catalog}

Some of the criticisms in NH13 arise from discrepancies between the 
text of S12 and the actual procedure used to generate our catalog.
Most importantly, we stated in S12 that we applied a maximum density 
threshold for each void. This was incorrect: we applied a minimum density 
criterion for merging adjacent zones into voids 
(as described in~\citealt{Neyrinck2008}). This does not limit 
the overall density of the voids, and if a void contains only a 
single zone it does not restrict the inclusion of that void.

Secondly, we misstated the handling of edge galaxies: we did not 
remove any galaxies, but followed a similar procedure as described in NH13 
and removed their adjacencies from the Voronoi graph, 
preventing {\tt ZOBOV} from including them in the watershed. Also, 
we considered any galaxy with any adjacency to a boundary particle as 
an ``edge'' galaxy, not just those that were closer to a boundary 
particle than any other galaxy.
Finally, in the original catalog the central density cut 
was erroneously applied at fixed radius, rather than at $0.25 R_{\rm eff}$.

We have also made several improvements, including the 
addition of void 
shape estimation and tools to extract void member particles.
We have documented all known bug fixes and improvements 
at {http://www.cosmicvoids.net} and in the catalog {\tt README}.

\section{Differences between S12 and NH13}

While the approach of NH13 is nearly identical to ours 
(e.g., the division into volume-limited samples, 
the use of {\tt ZOBOV}, and the procedure for handling boundaries),
they do introduce 
some innovations. Most importantly, they reject voids with 
$\rho_{\rm min} > 0.2 \bar{\rho}$, account for slight redshift-dependence 
in the mean galaxy density, include a bright-star mask, 
and apply different criteria for 
merging zones together.
We are
currently investigating the merits or demerits of such adjustments to
assess whether we should include them in future versions of our
catalog.

However, our fundamental 
philosophy is to minimally interfere with the void finding 
operation itself, produce a catalog with as many voids as possible, and 
allow users to apply post-processing filters and cuts as they 
see fit --- for example, users can already apply a $\rho_{\rm min}$ 
cut with the information we provide. The primary criticisms of 
NH13 rest on our choice to include the smallest voids 
in our catalog, and 
this study serves as a
reminder that small voids are more subject to Poisson noise and
statistically have different properties than larger voids.

\section*{Acknowledgements}

We thank S. Nadathur and S. Hotchkiss
 for sending an advance 
version of their paper and responding to our questions.  This
has led to improvements in our documentation of the void catalog,
including those contained in this response, even though we disagree
with them on the significance of a number of their points.

PMS and BDW acknowledge
support from NSF Grant AST-0908902. BDW
acknowledges funding from an ANR Chaire d’Excellence,
the UPMC Chaire Internationale in Theoretical Cosmology, and NSF grants AST-0908902 and AST-0708849. 
GL acknowledges support from CITA National Fellowship and financial
support from the Government of Canada Post-Doctoral Research Fellowship.
Research at Perimeter Institute is supported by the Government of Canada
through Industry Canada
 and by the Province of Ontario through the Ministry of Research and
Innovation.
DW acknowledges support from NSF Grant AST-1009505.


\footnotesize{
  \bibliographystyle{mn2e}
  \bibliography{refs}

\begin{thebibliography}{}

\bibitem[\protect\citeauthoryear{{Ceccarelli}, {Paz}, {Lares}, {Padilla} \&
  {Garc{\'{\i}}a Lambas}}{{Ceccarelli} et~al.}{2013}]{Ceccarelli2013}
{Ceccarelli} L.,  {Paz} D.,  {Lares} M.,  {Padilla} N.,    {Garc{\'{\i}}a
  Lambas} D.,  2013, ArXiv e-prints: 1306.5798

\bibitem[\protect\citeauthoryear{{Hamaus}, {Wandelt}, {Sutter}, {Lavaux} \&
  {Warren}}{{Hamaus} et~al.}{2013}]{Hamaus2013}
{Hamaus} N.,  {Wandelt} B.~D.,  {Sutter} P.~M.,  {Lavaux} G.,    {Warren}
  M.~S.,  2013, ArXiv e-prints: 1307.2571

\bibitem[\protect\citeauthoryear{{Melchior}, {Sutter}, {Sheldon}, {Krause} \&
  {Wandelt}}{{Melchior} et~al.}{2013}]{Melchior2013}
{Melchior} P.,  {Sutter} P.~M.,  {Sheldon} E.~S.,  {Krause} E.,    {Wandelt}
  B.~D.,  2013, ArXiv e-prints: 1309.2045

\bibitem[\protect\citeauthoryear{{Nadathur} \& {Hotchkiss}}{{Nadathur} \&
  {Hotchkiss}}{2013}]{Nadathur2013}
{Nadathur} S.,  {Hotchkiss} S.,  2013, ArXiv e-prints: 1310.2791

\bibitem[\protect\citeauthoryear{Neyrinck}{Neyrinck}{2008}]{Neyrinck2008}
Neyrinck M.~C.,  2008, \mnras, 386, 2101

\bibitem[\protect\citeauthoryear{{Pisani}, {Lavaux}, {Sutter} \&
  {Wandelt}}{{Pisani} et~al.}{2013}]{Pisani2013}
{Pisani} A.,  {Lavaux} G.,  {Sutter} P.~M.,    {Wandelt} B.~D.,  2013, ArXiv
  e-prints: 1306.3052

\bibitem[\protect\citeauthoryear{{Planck Collaboration}}{{Planck
  Collaboration}}{2013}]{Planck2013b}
{Planck Collaboration} 2013, ArXiv e-prints: 1303.5079

\bibitem[\protect\citeauthoryear{{Ryden} \& {Melott}}{{Ryden} \&
  {Melott}}{1996}]{Ryden1996}
{Ryden} B.~S.,  {Melott} A.~L.,  1996, \apj, 470, 160

\bibitem[\protect\citeauthoryear{{Sutter}, {Lavaux}, {Wandelt}, {Hamaus},
  {Weinberg} \& {Warren}}{{Sutter} et~al.}{2013}]{Sutter2013a}
{Sutter} P.~M.,  {Lavaux} G.,  {Wandelt} B.~D.,  {Hamaus} N.,  {Weinberg}
  D.~H.,    {Warren} M.~S.,  2013, ArXiv e-prints: 1309.5087

\bibitem[\protect\citeauthoryear{{Sutter}, {Lavaux}, {Wandelt} \&
  {Weinberg}}{{Sutter} et~al.}{2012}]{Sutter2012b}
{Sutter} P.~M.,  {Lavaux} G.,  {Wandelt} B.~D.,    {Weinberg} D.~H.,  2012,
  \apj, 761, 187

\bibitem[\protect\citeauthoryear{{van de Weygaert} \& {Platen}}{{van de
  Weygaert} \& {Platen}}{2011}]{VandeWey2011b}
{van de Weygaert} R.,  {Platen} E.,  2011, International Journal of Modern
  Physics Conference Series, 1, 41

\end{thebibliography}
}

\end{document}